\documentclass{emulateapj}
\usepackage{apjfonts}

\def\ltsima{$\; \buildrel < \over \sim \;$}
\def\simlt{\lower.5ex\hbox{\ltsima}}
\def\gtsima{$\; \buildrel > \over \sim \;$}
\def\simgt{\lower.5ex\hbox{\gtsima}}

\slugcomment{}

\shorttitle{GWs from Newborn Magnetars}
\shortauthors{Stella et al.}

\begin{document}

\title{Gravitational Radiation from Newborn Magnetars in the Virgo Cluster} 

\author{L. Stella, S. Dall'Osso, G. L. Israel}
\affil{INAF - Osservatorio Astronomico di Roma, \\ Via Frascati 33, 
  00040 Monteporzio Catone, Italy.}
\email{stella, dallosso, gianluca@mporzio.astro.it}

\and

\author{A. Vecchio}
\affil{School of Physics and Astronomy, University of Birmingham, \\ Edgbaston,
 Birmingham B15 2TT, UK}
\email{av@star.sr.bham.ac.uk}

\begin{abstract}

There is growing evidence that two classes of high-energy 
sources, the Soft Gamma Repeaters and the 
Anomalous X-ray Pulsars  contain slowly spinning 
``magnetars'', {\it i.e.} neutron stars whose emission is powered 
by the release of energy from their extremely strong magnetic fields
($>10^{15}$~G).
We show here that the enormous energy liberated in the 2004 December 27 giant 
flare from SGR~1806-20 ($\sim 5 \times 10^{46}$~erg), together with 
the likely recurrence time of such events, requires an internal field 
strength of $ \simgt 10^{16}$~G.  
Toroidal magnetic fields of this strength are within an order of magnitude 
of the maximum fields that can be generated in the 
core of differentially-rotating neutron stars immediately after their 
formation, if their initial spin period is of a few milliseconds. 
A substantial deformation of the neutron star is induced by these magnetic 
fields and, provided the deformation axis is offset from the spin axis, a 
newborn fast-spinning magnetar would radiate for a few weeks 
a strong gravitational wave signal the frequency of which 
($\sim 0.5-2$~kHz range) decreases in time. 
The signal from a newborn magnetar with internal field  
$> 10^{16.5}$~G could be detected with Advanced LIGO-class 
detectors up to the distance of the Virgo cluster 
(characteristic amplitude $h_{c}\sim 10^{-21}$).  
Magnetars are expected to form in Virgo at a rate $\simgt 1$~yr$^{-1}$.
If a fraction of these have sufficiently high 
internal magnetic field, then newborn magnetars constitute
a promising new class of gravitational wave emitters. 

\end{abstract}


\section{Introduction}

The two classes of pulsating high energy sources that are 
believed to host magnetars, the Soft Gamma  Repeaters, SGRs, and the 
Anomalous X-ray Pulsars, AXPs, share a number of properties (see e.g. 
\citet{MeSte95,Kou98,WooTho04}). 
They have fairly long spin periods in the $\sim 5\div10$~s range, spin-down 
secularly with timescales $\sim 10^4\div10^5$~yr, 
do not have a companion star and are in some cases associated to 
supernova remnants with ages $\sim 10^3\div10^4$~yr. Rotational energy losses 
are $1\div2$ orders of magnitude too low to explain the persistent emission 
of these sources, which is typically $\sim 10^{34}\div10^{35}$ erg/s. 
Both AXPs 
and SGRs show periods of activity in which recurrent short bursts ($\ll 1$~s) 
are emitted, with peak luminosities in the $\sim 10^{38}-10^{41}$~erg/s range. 
The initial spikes of giant flares are second-long events of exceptionally high
luminosity, 3 to 6 orders of magnitude larger than that of the brighest 
recurrent bursts. Giant flares are rare, only three have been observed so far, 
and might well be specific to SGRs. Given the highly super-Eddington 
luminosities of recurrent bursts and, especially, giant flares, accretion power
is not an option.

The magnetar model envisages that SGRs and AXPs are powered by 
the release of energy from their extremely high magnetic 
fields \citep{DT92,TD93,TD95,TD96,TD2001}. The model has been largely 
successful in interpreting the unique features of these sources.
According to it, a twisted, mainly toroidal magnetic field characterizes  
the neutron star interior ($B>10^{15}$~G), only partially threading the crust. 
The emerged (mainly poloidal) field makes up the neutron star magnetosphere, 
with dipole strenghts of $B_d \sim $ few $\times 10^{14}$~G, 
as required to generate spin-down at the 
observed rate \citep{TD93,TM01}. Energy is fed to the neutron star 
magnetosphere through Alfv\'{e}n waves driven by local 
``crustquakes" of various amplitude and producing recurrent bursts. 
Giant flares may result from large-scale 
rearrangements of the core magnetic field or catastrophic 
instabilities in the magnetosphere \citep{TD2001,Lyu03}.
These events occur more rarely and lead 
to the sudden release of very large amounts of energy. 
Most of this energy breaks out of the magnetosphere in the form of a fireball 
of plasma expanding at relativistic speeds and giving rise to 
the initial spike of giant flares. The oscillating tail that follows this 
spike, displaying up to $\sim 50 $ cycles of the neutron star spin, is 
interpreted as due to that part of the flare energy that 
remains trapped in the magnetosphere. The total energy released in this tail 
($\sim 10^{44}$~erg in all three events detected so far) provides an 
independent estimate of $\simgt 10^{14}$~G for the external field of magnetars 
\citep{TD95,TD2001}.

\section{The Internal B-field of  Magnetars}

The 2004 December 27 giant flare from SGR1806-20 provides a new estimate of the
internal field of magnetars.
About $ 5 \times 10^{46}$~erg were released in the $\sim 0.6$~s long main 
spike of this event \citep{Teras05,Hur05} {\it i.e.} a two decade higher 
energy than that of the other giant flares observed so far, the 1979 
March 5 event from SGR 0526-66 \citep{Maz79} and the 1998 August 27 event from 
SGR 1900+14 \citep{Hur99,Fer99}.
Only one such powerful flare has been recorded in about 30 yr of monitoring
from the $\sim 5$ known magnetars in SGRs.  The recurrence time in a single 
magnetar implied by this event is thus about $\sim 150 $~yr (if giant flares 
were emitted also by AXPs, this recurrence time would increase by a factor of 
$\sim 2$, as the number of AXPs and SGRs in our galaxy is comparable). The 
realisation that powerful giant flares could be observed from distances of tens
of Mpc (and thus might represent a sizeable fraction of the short Gamma Ray 
Burst population) motivated searches for 2004 Dec 27-like events in the 
BATSE GRB database \citep{Laz05,Popo05}. The upper limits on the recurrence 
time of powerful giant flares obtained in these studies range from $\tau \sim 
130$ to 600~yr per galaxy, {\it i.e.} $\sim 4$ to 20 times longer than the time
inferred from the 2004 Dec 27 event. A detailed discussion of the merits of 
these estimates is beyond the scope of this Letter. However, we note 
that a 2004 Dec 27-like event in the Galaxy could not be missed, whereas 
several systematic effects can reduce the chances of detection from 
large distances (see e.g. the discussion in \citet{Laz05,Nak05}). 
Rather than regarding the 
2004 Dec 27 event as statistically unlikely, one can evaluate the 
chances of a recurrence time of hundreds of years, given the 
occurrence of the 2004 Dec 27 hyperflare. 
We estimate that, having observed a powerful giant flare in our galaxy in 
$\sim 30$~yr of observations, the Bayesian probability that the 
galactic recurrence time is $\tau > 600$~yr is $\sim 10^{-3}$, 
whereas the $90\%$ confidence upper limit is $\tau \sim 60$~yr.
These valeus are derived for a uniform prior\footnote{For a uniform prior 
on $\log (1/\tau)$ and a (conservative) lower cutoff of $1/(10^4 yr)$, 
we obtain $\tau > 600$~yr with probability $0.04$ and 
a $90\%$ confidence upper limit of $\tau \sim 270$~yr.}
on the rate $1/\tau$.  
We thus favor smaller values and assume in the following $\tau \sim 30$~yr. 

In a timescale of 
$\sim 10^4$~yr (that we adopt for the SGR lifetime) about 70 very powerful 
giant flares should be emitted which release a total energy of $\sim 4 \times 
10^{48}$~erg. 
We note that if the powerful giant flares' emission were beamed in a fraction 
$b$ of the sky (and thus the energy released in individual flares a factor of 
$b$ lower), the recurrence time would be a factor of $b$ shorter. Therefore the
total energy released by giant flares would remain the same. 
For this energy to originate in the magnetar's B-field, this must be 
$\geq 10^{15.7}$~G. This value should be regarded as a {\it lower 
limit} on the initial internal field of a magnetar. 
Firstly, not all the the magnetic energy is released 
through powerful giant flares. The magnetar model predicts 
a conspicuous neutrino luminosity, resulting from ambipolar diffusion, of  
$L_{\nu} \sim 4\times 10^{36}~t^{-8/7}_4$~erg~s$^{-1}$, 
where $t_4$ is the magnetar age in units of $10^4$ yr \citep{TD96}. 
Integrating over $t_4 = (0.01 \div 1)$ yields a neutrino energy output 
of $\sim 4 \times 10^{48}$~erg. Including this energy gives an 
internal magnetic field of $\geq 10^{15.9}$ G. 
Secondly, ambipolar diffusion and magnetic dissipation are expected to 
take place at a faster rate for higher values of the field 
\citep{TD96}. Therefore estimates 
of the B-field based on present day properties of SGRs (age $\sim 10^4$~yr) 
likely underestimate the value of their initial magnetic field. 

Very strong internal B-fields are expected to be generated in a 
differentially rotating fast spinning neutron star, subject to vigorous 
neutrino-driven convection instants after its formation \citep{DT92}. 
If the initial spin period $P_i$ is in $\sim 1\div2$~ms range  
(comparable to the overturn time of convective cells), an efficient 
$\alpha-\Omega$ dynamo is powered by fluid helycal motions, which amplifies the
field in a predominantly toroidal configuration with a coherence length 
comparable to the star size. Differential rotation 
provides a very large free energy reservoir, $\sim 10^{52}~(P_i/\mbox{1 ms})^
{-2}$~erg, that can be converted into a magnetic field of up to 
$\sim 3 \times 10^{17} (P_i/1~\mbox{ms})^{-1}$ G \citep{DT92}.
However the efficiency of the amplification is likely limited by the dynamical 
influence (backreaction) of the field which opposes convective motions.
This limitation could be circumvented, however, if smoothing of 
angular velocity gradients by magnetic stresses is very efficient on a 
timescale shorter than the initial neutrino cooling time (which sets the 
timescale for convective instability to occur); in this case, a field 
of several $\times 10^{16}$ G can be generated in magnetars that are born with 
spin periods of a few milliseconds \citep{TD93,Dun98}.

In summary, unless powerful giant flares such as the 2004 December 27 event are
much rarer than the rate implied by having already detected one of them, 
magnetars must possess internal fields of $\sim 10^{16}$~G or higher  
(values up to $\sim 10^{17}$~G cannot be ruled out).
Based on current models, fields of this order can be generated only if 
magnetars are born with spin periods of a few ms. 
In the following we explore the consequences of this for the generation of 
gravitational waves from newborn magnetars. 
We parametrize their (internal) toroidal field with $B_{t,16.3} = B_t/
2\times 10^{16}$~G, (external) dipole field with $B_{d,14} = 
B_d/10^{14}$~G and initial spin period with $P_{i,2} = 
P_i/(2~\mbox{ms})$. 

\section{Magnetically-Induced Distortion and Gravitational Wave Emission}

The possibility that fast-rotating, magnetically-distorted neutron stars are
conspicuous sources of gravitational radiation has been discussed by several
authors (see e.g. the review by \citet{Bona94}). The path towards a better 
understanding of the early evolution and gravitational radiation emission 
properties of such neutron stars was layed out by \citet{Cut02}, who 
discussed also the possibility of detecting the gravitational radiation 
signal from a newborn, $\sim 10^{15}$~G neutron star in our Galaxy. If this 
value for the internal field pertains to ordinary neutron stars, 
then such events might be expected at the galactic rate of neutron star 
formation, {\it i.e.} once every several tens to hundreds of years 
(see e.g. \citet{Arz02}). 
The formation rate of magnetars is substantially lower than that (see 
Section 4), 
making the chances of witnessing the birth of a galactic magnetar in the 
near future very slim. In the following we show that for the range of 
magnetic fields discussed in Section 2, newly born, millisecond spinning 
magnetars are conspicuous sources of gravitational radiation that might be 
detectable up to the distances of the Virgo cluster.

The anisotropic pressure from the toroidal B-field deforms a magnetar
into a prolate shape, with ellipticity $\epsilon_B \sim -6.4 \times
10^{-4} (<B^2_{t,16.3}>)$, where the brackets indicate a volume-average over 
the entire core \citep{Cut02}.
In general, the symmetry axis of the magnetically-induced deformation will not 
be coaligned with the rotation axis.
Hence, in the neutron star frame the angular 
velocity vector undergoes free precession with period $P_{prec} \simeq P / 
\epsilon_B$ around the (magnetic) symmetry  axis, where $P$ is the spin period 
\citep{Jon02,JAnd02}
(the rotationally-induced distortion is not relevant here, because it is 
always aligned with the instantaneous spin axis).
As long as the axis of the magnetic distortion is not aligned with the 
spin axis, the star's 
rotation will cause a periodic variation of the mass quadrupole moment, in turn
resulting in the emission of gravitational waves, GWs, at twice the spin 
frequency of the star. 

A crucial issue is whether the misalignement of the magnetic axis can be 
maintained in spite of the damping of free precession caused by the star's 
internal viscosity. 
In a prolate ellipsoid the rotational energy is minimized when the moment 
of inertia along the axis of the ``frozen-in'' distortion, $I_3$, is 
orthogonal to the spin axis \citep{JAnd02,Cut02}. 
Hence, viscous damping of precession drives a 
prolate neutron star towards a geometry that maximizes the time-varying mass 
quadrupole moment and, thus, GW emission.
 
As the power emitted in GWs scales as $\propto P^{-6}$, the GW signal, for a 
given internal B-field, depends critically on the initial spin period and its 
variation thereafter. The spin evolution of a newborn magnetar is determined by
angular momentum losses from GWs, electromagnetic dipole radiation and 
relativistic winds. According to \citet{ThChQu04},
the latter mechanism depends critically on the value of 
the external dipole and is negligible except for fields 
$> (6\div 7) \times 10^{14}$~G, a regime that we do not discuss
here. 
We consider in the following the spin evolution of a newborn magnetar under the
combined effects of GW and electromagnetic dipole radiation. 
When both mechanisms are taken into account, the magnetar 
angular velocity $\omega=2\pi/P$ evolves according to 
\begin{equation}
\label{general}
\dot{\omega} = - \frac{K_d}{2}\omega^3 - \frac{K_{GW}}{4} \omega^5 \ ,
\end{equation}
where $K_d = (B^2_d R^6)/(3 I c^3)$ 
and $K_{GW} = (128/5) (G/c^5) I \epsilon^2_B 
$, with $R$ the neutron star radius, $G$ the gravitational constant and $c$ the
speed of light. From this equation, we derive a spin-down timescale of
\begin{equation}
\label{tausd}
\tau_{sd} \equiv \omega/(2 \dot{\omega}) 
\simeq 10~~P^2_{i,2}\left(B_{d,14}^2 + 1.15 B_{t,16.3}^{4}
P_{i,2}^{-2} \right)^{-1} \mbox{d} \ .
\end{equation}
According to models of neutron star internal viscosity \citep{APin85,Alp88}, 
orthogonalization of the magnetic axis is expected to take place in $n\simeq 
10^4$ precession cycles \citep{Cut02}. This translates into a timescale of
$\tau_{ort} =  n P_i/\epsilon_B \sim 0.4 B^{-2}_{t,16.3} P_{i,2}$~d. 
The condition for the newly formed magnetar to become
an orthogonal rotator before loosing most of its spin energy is 
\begin{equation}
\label{condition}
\frac{\tau_{sd}}{\tau_{ort}} \simeq 26~\frac{ B^2_{t,16.3}}{B^2_{d,14} P^{-1}_
{i,2} + 1.15~B^4_{t,16.3} P^{-3}_{i,2}} > 1 \ .
\end{equation}
If condition (\ref{condition}) is met, the magnetar quickly becomes a maximally
efficient GW emitter, while its spin period is still close to the initial one. 
The strain is 
\begin{equation}
\label{strain}
h \sim 3 \times 10^{-26} d_{20}^{-1} P^{-2}_{2} B^2_{t,16.3}\ ,
\end{equation} 
where the distance $d_{20}=d/(20 \mbox{ Mpc})$  
is in units of the Virgo Cluster distance and the angle-averaged strain 
is that given by \citet{UshCutBil00}.      
We estimate the characteristic amplitude, $ h_{c} = h N^{1/2}$, 
where $N\simeq \tau_{sd}/P_i$ is the 
number of cycles over which the signal is observed. 
Using equation (\ref{strain}) we obtain
\begin{equation} 
\label{effective}
h_{c} \simeq 6 \times 10^{-22} d_{20}^{-1}
B_{t,16.3}^2 P^{-3/2}_{i,2}\left(B_{d,14}^2 + 1.15 B_{t,16.3}^4 
P_{i,2}^{-2}\right)^{-\frac{1}{2}}  \ .
\end{equation}
Under the conditions discussed above, strong GW losses are not quenched 
immediately after the magnetar birth but rather extend in time, typically from 
days to a few weeks, before fading away as a result of the star spin-down.

\begin{figure}
\plotone{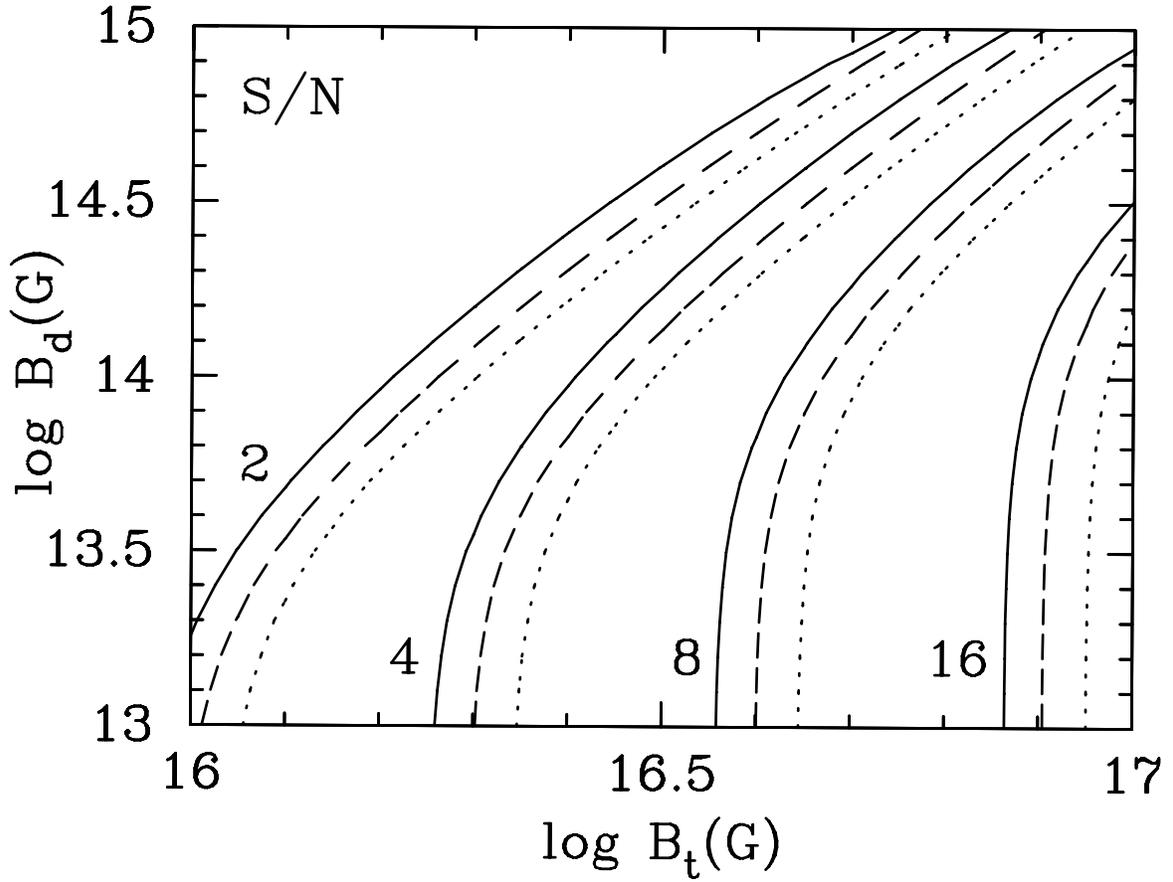}
\caption{Lines of constant $S/N$ for selected values of the 
initial spin period in the internal toroidal magnetic field, $B_t$, 
and external dipole field, $B_t$, plane for a source 
at the distance of the Virgo Cluster ($d_{20}=1$). Solid, dashed and dotted 
curves correspond to an initial spin period of $P_i=$1.2, 2 and 2.5~ms,
respectively. The calculations take into account the time required for 
the toroidal magnetic field axis to become orthogonal to the spin axis.  
Note that according to \citet{ThChQu04}, 
strong angular momentum losses by relativistic 
winds set in and dominate the spin down for $B_d > 6\div7 \times 10^{14}$~G;  
the curves for such values of $B_d$ should thus be treated with caution.}
\label{gwsall}
\end{figure}

In order to assess the detectability of this GW signal we
compute the optimal (matched-filter) signal-to-noise ratio
\begin{equation}
\label{total}
\left(\frac{S}{N}\right) = 2
\left(\int_{f_f}^{f_i} \frac{|\tilde{h}(f)|^2}
{S_h(f)}~ df\right)^
{\frac{1}{2}} 
\end{equation}
by extending the formulae by \citet{OwLind02} and \citet{Cut02} 
to include both the electromagnetic and GW torques. In the previous
equation $f_i = 2/P_i \approx 2$ kHz and $f_f = 2/P_f\approx 500$ Hz are
the initial and final values of the frequency of the signal. $\tilde h(f)$
is the Fourier transform of the GW strain at the detector 
(Eq. (\ref{strain}) divided by $5^{1/2}$ in order to account for averaging 
over source sky position) and $S_h(f)$
the one-sided noise spectral density of the instrument. In the
relevant frequency band 500 Hz - 2 kHz we approximate
$S_h(f)$ according to the current baseline performance of Advanced LIGO:
$S_h (f) = 2.1\times 10^{-47} (f/1\,\mathrm{kHz})^2\,\mathrm{Hz}^{-1}$ 
\citep{Tho00}.
We compute $\tilde h(f)$ using the stationary phase approximation 
(see e.g. \citet{OwLind02}) with
angle-averaged amplitude and frequency derivative $\dot{f} = \dot{\omega}/\pi$ 
given by Eqs. (\ref{strain}) and (\ref{general}), respectively. 
Eq.(\ref{total}) yields the following estimate of the optimal signal-to-noise
ratio:
\begin{equation}
\label{soluzione}
\left(\frac{S}{N}\right) \simeq 3 d_{20}^{-1}
B^2_{t,16.3}B^{-1}_{d,14} \left[\mbox{ln}\left(\frac{a^2+ f_f^2}{a^2+ f_i^2}
\right)+ 2 \mbox{ln}\left(\frac{f_i}{f_f}\right)\right]^{\frac{1}{2}} \ ,
\end{equation}
where $a^2 \equiv 2 K_d/(\pi^2 K_{GW})$.
The limit in which the electromagnetic torque dominates corresponds to 
$ a^2 \gg f^2$ (see Eq. (\ref{general})). In this case, the first logarithmic 
term  in Eq. (\ref{soluzione}) is negligible and we recover the 
result by \citet{Cut02}. Fig.~\ref{gwsall} shows lines of
constant $S/N$ for a source at $d_{20} = 1$  and selected
values of the initial rotation period ($P_i = 1.2,2$ and 2.5~ms) in
the $(B_t,B_d)$ plane: GWs from newborn magnetars can produce $S/N > 8$ 
for $B_t\ge 10^{16.5}$ G and 
$B_d \le 10^{14.5}$ G. This is the region of parameter space that offers
the best prospects for detection as we now discuss.

The optimal (in the maximum likelihood sense) detection statistic 
for long-lived signals is the
so-called ${\cal F}$-statistic \citep{Jara98}; its computation
involves the correlation of the data stream with a discrete set of filters
that probe the relevant space of unknown parameters: the 
sky position (2 parameters) and 2 parameters that control the evolution
of the GW phase (see Eq.(\ref{general})). A very conservative upper-limit on 
the required number of filters involved in a search can be evaluated 
ignoring correlations among the parameters: $\sim 100$ filters 
\citep{Brady98} are needed to cover the $\sim 40\,\mathrm{deg}^2$ 
area of the Virgo cluster for a coherent integration time $T\sim 10^6$~s,  
and $\sim (T\,f_i)^2 \sim 10^{18}$ filters 
for the phase parameters, giving a total of $\sim 10^{20}$ templates. For
this number of trials, we estimate that a signal with $S/N \sim 12$ 
can be detected with  1\% false alarm and 10\% false dismissal 
(if all the parameters of the signal were known the corresponding
$S/N$ would be $\approx 4.6$). Of course the actual sensitivity limit of
such a search is likely set by the available computational resources. In
practice, the search for GWs from newborn magnetars is probably
best carried out using a ``hierarchical search strategy'' where 
coherent and incoherent stages are alternated in order to achieve 
(quasi-)optimal sensitivity at affordable computational costs
\citep{Brady00,Kri04,Cut05,Abb05}.
Furthermore, the presence of a trigger based on the detection of the
newborn magnetar by other means (e.g. the corresponding supernova)
could further reduce the region of parameter space that needs 
to be searched. In addition the sensitivity will increase with 
the operation of the network of GW detectors that is 
coming on line. In any case the search for this new class of signals 
represents a challenge that we are investigating in greater depth. 
\section{Discussion}
We have shown that the energy liberated in the 2004 December 27 flare from 
SGR~1806-20, together with the likely recurrence rate of these events, points 
to a magnetar internal field strength of $\sim 10^{16}$~G or greater.
Such a field is sufficient to induce a sizeable deformation of the star and is
most likely generated in a differentially rotating, millisecond spinning 
magnetar instants after gravitational collapse. 
Magnetars with these characteristics are thus expected to be very powerful 
sources of gravitational radiation during the first weeks of their life. 
Prospects for revealing their GW signal depend critically on the birth rate 
of these objects. 
The three confirmed associations between an AXP and a supernova remnant 
(ages in the $10^3 \div 10^4$~yr range) implies a magnetar birth rate of 
$\simgt 0.5\times 10^{-3}$~yr$^{-1}$ in the Galaxy \citep{Gae99}. 
The chances of revealing the signal from a 
newborn magnetar in our Galaxy are thus very low. 
A rich cluster like Virgo, containing $\sim$ 2000 galaxies, is expected to 
give birth to magnetars at a rate of $\simgt$ 1 yr$^{-1}$. A fraction of these
magnetars might have sufficiently high toroidal fields that a detectable GW 
is produced. A slowly 
evolving periodic GW signal at $\sim 1$~kHz, whose frequency halves 
over weeks, would unambiguously reveal the early days of fast spinning 
magnetars.
We conclude that newborn, fast spinning magnetars provide a class of GW 
emitters over Virgo scale distances that might well be within reach  
for the forthcoming generation of GW detectors. 

\acknowledgments

We acknowledge useful comments by the referee. 
This work was partially supported through ASI and MIUR grants.  
 


\end{document}